\documentclass[preprint,jcp,floats,amsmath,amssymb]{revtex4}

\usepackage{graphicx}
\usepackage{amsmath}
\usepackage{amssymb}
\usepackage{subfigure}
\usepackage[english]{babel}
\usepackage{MnSymbol}
\usepackage{color}
\usepackage{chemarr}

\begin{document}
\title{The linear-noise approximation and the chemical master equation exactly agree up to second-order moments for \\ a class of chemical systems}
\author{Ramon Grima $^{1}$}
\affiliation{$^{1}$ School of Biological Sciences, University of Edinburgh, UK}

\begin{abstract}
It is well known that the linear-noise approximation (LNA) exactly agrees with the chemical master equation, up to second-order moments, for chemical systems composed of zero and first-order reactions. Here we show that this is also a property of the LNA for a subset of chemical systems with second-order reactions. This agreement is independent of the number of interacting molecules. 
\end{abstract}

\maketitle

\section{Introduction}

The Markovian description of chemical systems, as given by the chemical master equation (CME) \cite{McQuarrie1967}, is rarely exactly solvable. This has led to the development of a number of approximation techniques such as moment-closure approximations \cite{Verghese2007,Ale2013} and the linear-noise approximation (LNA) \cite{vanKampen,vanKampen1961}. The accuracy of these techniques is frequently unknown and hence there is a substantial ongoing effort to clearly understand the limitations of the techniques and the magnitude of the error inherent in their predictions (see for example \cite{Ferm2008,Grima2010,Grima2011,GrimaJCP2012,Wallace2012}). 

Here we exclusively focus on the LNA. The LNA originates from the system-size expansion first introduced by van Kampen \cite{vanKampen1961}. The system-size expansion effectively constitutes an infinite series expansion of the moments of the probability distribution solution of the CME in powers of the inverse volume of the compartment in which the chemical system is confined. The LNA is the leading-order term in this expansion which implies that in this approximation, the mean concentrations are the same as given by the conventional rate equations while the variance of concentration fluctuations is proportional to the inverse volume. Hence the LNA is conventionally regarded as a large volume (macroscopic) approximation of the moments of the CME; equivalently the LNA an be viewed as an accurate approximation in the limit of large molecule numbers, since the macroscopic limit is the limit of large volumes at constant concentration \cite{vanKampen}. Remarkably, however, the LNA is exact (up to second-order moments) for systems composed of zero and first-order reactions. This is since in this case the propensities (the transition rates in the CME) are already linear in the molecule numbers and hence the linearisation procedure inherent in the application of the LNA bears no effect on the equations for the time-evolution of the covariance matrix (see for example \cite{Lestas2008}). It has also been shown that the differences between the predictions of the rate equations / LNA and of the CME are proportional to the elements of the Hessian matrix of the rate equations \cite{Grima2010,Grima2011,Thomas2014} and inversely proportional to the volume. Since the Hessian matrix is non-zero whenever chemical systems have at least one second-order reaction, it is generally thought that the LNA's predictions increase with the non-linear dependence in the molecule numbers in the propensities of the CME and with decreasing the volume, a claim supported by several case studies \cite{Ferm2008,Scott2012,Thomas2013,Hayot2004,Cianci2012}.

Summarising, the current general picture is that (i) the LNA is exact (up to second-order moments) for systems composed of at most first-order reactions since the propensities are constant or directly proportional to the molecule numbers. (ii) the LNA is inexact for systems with at least one second-order reaction since some of the propensities are proportional to the product of molecule numbers.

In this article, we show that this standard picture needs revision. In particular, we prove that there exists a special class of chemical systems containing at least one second-order reaction for which the error in LNA's prediction of the mean concentrations and of the variance of fluctuations about the means is zero for all volumes and parameters of the system. This implies that for these systems, the LNA is exact for all molecule numbers and not just accurate in the limit of large molecule numbers, as commonly thought.  

The article is organised as follows. In Section II, we show by means of examples, whose CME can be solved exactly, that the rate equations and LNA for systems with at least a second-order reaction, can in some cases lead to exact expressions for the mean concentrations and the variance of the fluctuations about them. In Section III, we identify a broad class of chemical systems with this property.  In Section IV, we generalise further this special class of systems and provide several examples of common chemical systems of this type. We finish by a discussion in Section V. 

\section{When are the rate equations and the LNA exact?}

\subsection{An illustrative example}

Consider the following two different reaction systems:
\begin{align}
\label{r1ex}
&X_1 + X_2 \xrightleftharpoons[k_1]{k_0} X_3, \\
\label{r2ex}
&X_1 + X_2 \xrightleftharpoons[k_1]{k_0} X_3, \ \O  \xrightleftharpoons[k_3]{k_2} X_1,
\end{align}
where the rate constants are those which would appear in a rate equation formulation of the systems. Reaction (\ref{r1ex}) is a closed heterodimerisation reaction whereby molecules of species $X_1$ and $X_2$ reversibly combine to form a heterodimer $X_3$. This system has two implicit conservation laws: $n_2 - n_1 = \alpha$ and $n_1 + n_3 = \beta$ where $n_i$ is the number of molecules of species $X_i$ and $\alpha, \beta$ are time-independent constants. The first conservation law is due to the fact that whenever a molecule of $X_1$ is produced (or consumed), a molecule of $X_2$ is also produced (or consumed); the second conservation law stems from the fact that whenever a molecule of $X_1$ is produced (or consumed), a molecule of $X_3$ is consumed (or produced). Reaction (\ref{r2ex}) is an open version of the previous reaction since molecules of $X_1$ are also produced and destroyed. There is one implicit conservation law in this system, namely $n_2 + n_3 = \gamma$ where $\gamma$ is a time-independent constant. 

In steady-state conditions, both systems satisfy detailed balance and hence using standard methods \cite{vanKampen1976} the exact solution of the CME's of both systems can be straightforwardly obtained. These are given by:
\begin{align}
\label{Psys1}
P(n_1) &= \frac{\alpha! \beta! (\frac{k_1 \Omega}{k_0})^{n_1}}{n_1! (n_1 + \alpha)! (\beta - n_1)! M(-\beta,1+\alpha,-\frac{k_1 \Omega}{k_0})}, \\
\label{Psys2}
P(n_1,n_2) &= e^{-\frac{k_2 \Omega}{k_3}} \frac{(\frac{k_2 \Omega}{k_3})^{n_1}}{n_1!} \biggl(1 + \frac{k_3 k_1}{k_2 k_0} \biggr)^{-\gamma} \gamma!  \frac{(\frac{k_3 k_1}{k_2 k_0})^{n_2}}{n_2! (\gamma - n_2)!},
\end{align}
for reaction systems (\ref{r1ex}) and (\ref{r2ex}) respectively. The volume of the compartment in which the reaction occurs is given by $\Omega$. Note that there is no explicit $n_2, n_3$ dependence in Eq. (\ref{Psys1}) since $n_1$ is related to $n_2,n_3$ via the implicit conservation law. Similarly there is no $n_3$ dependence in Eq. (\ref{Psys2}) since $n_2$ is related to $n_3$ via the implicit conservation law. The function $M(x,y,z)$ denotes the Kummer confluent hypergeometric function \cite{Abramowitz}. 

The rate equations for the chemical system (\ref{r1ex}) are given by:
\begin{align}
\frac{d}{dt} \phi_1 = \frac{d}{dt} \phi_2 = -\frac{d}{dt} \phi_3 = -k_0 \phi_1 \phi_2 + k_1 \phi_3,
\end{align}
while for the chemical system (\ref{r2ex}) they are given by:
\begin{align}
\frac{d}{dt} \phi_1 &= k_2 - k_3 \phi_1 - k_0 \phi_1 \phi_2 + k_1 \phi_3, \\
\frac{d}{dt} \phi_2 &= -\frac{d}{dt} \phi_3 = -k_0 \phi_1 \phi_2 + k_1 \phi_3,
\end{align}
where $\phi_i = \langle n_i \rangle / \Omega$ is the concentration of species $X_i$. These equations when solved in steady-state conditions, yield the following mean number of molecules:
\begin{align}
\label{REsys1}
\langle n_1 \rangle &=  \frac{-\alpha k_0 - k_1 \Omega + \sqrt{4 \beta k_0 k_1 \Omega + (\alpha k_0 + k_1 \Omega)^2}}{2 k_0}, \quad  \langle n_2 \rangle = \langle n_1 \rangle + \alpha, \quad \langle n_3 \rangle = \beta - \langle n_1 \rangle, \\ \langle n_1 \rangle &= \frac{k_2 \Omega}{k_3}, \quad \langle n_2 \rangle = \frac{k_1 k_3 \gamma}{k_0 k_2 + k_1 k_3},  \quad \langle n_3 \rangle = \gamma -  \langle n_2 \rangle.
\label{REsys2}
\end{align}
for the reaction systems (\ref{r1ex}) and (\ref{r2ex}) respectively. 

Computing the mean number of molecules from the exact probability distribution solution Eq. (\ref{Psys1}) of the CME for system (\ref{r1ex}), one finds that they are generally different from the solution of the rate equations given by Eq. (\ref{REsys1}). These differences are illustrated (by the blue and green lines) in Fig. 1(a) for the parameters $k_0 = 1, k_1 = 0.1, \alpha = 0, \beta = \Omega$ with the  volume $\Omega$ varied in discrete steps of $1$ such that the quantity $\beta$ is an integer (this is required by the conservation law discussed earlier). Note that $\alpha = 0$ and $\beta = \Omega$ imply that the concentrations of $X_1$ and $X_2$ are equal at all times and that sum of the concentrations of $X_1$ and $X_3$ is unity at all times, respectively. In contrast, the mean number of molecules computed from the exact probability distribution solution Eq. (\ref{Psys2})  of the CME for system (\ref{r2ex}), are exactly the same as those given by the rate equation solution Eq. (\ref{REsys2}) -- this is illustrated by the dashed red line in Fig. 1(a).  

Similarly one can show that the Fano factors for both species (the ratio of the variance of fluctuations and of the mean number of molecules) computed using the exact probability distribution solutions differ from those obtained using the LNA for system (\ref{r1ex}) (illustrated in Fig. 1(b) for the same parameters used in Fig. 1(a)) and agree exactly with those obtained from the LNA for system (\ref{r2ex}). 

As mentioned in the Introduction, generally it is thought that the rate equations and LNA are only exact (up to second-order moments) for systems composed of at most first-order reactions and that for systems with at least a second-order reaction, they agree with the CME only in the limit of large volumes. From this perspective, the exact agreement found for system (\ref{r2ex}), for all volumes, is perplexing since this system does possess a second-order reaction. Next we show that system (\ref{r2ex}) forms part of a broad class of chemical systems for which the LNA is exact up to second-order moments.


\begin{figure}
\centering
\subfigure[]{
\includegraphics[width=85mm]{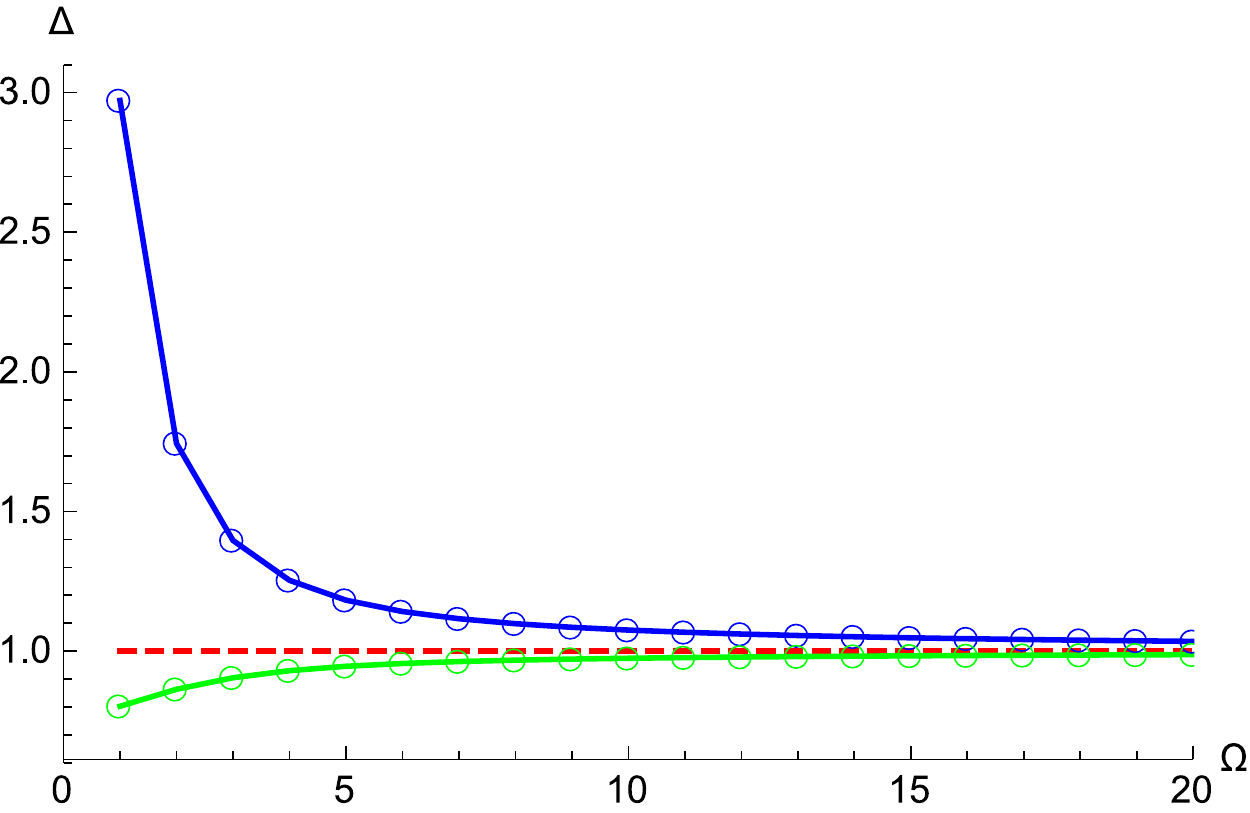}
\label{fig:subfig1}
}
\subfigure[]{
\includegraphics[width=85mm]{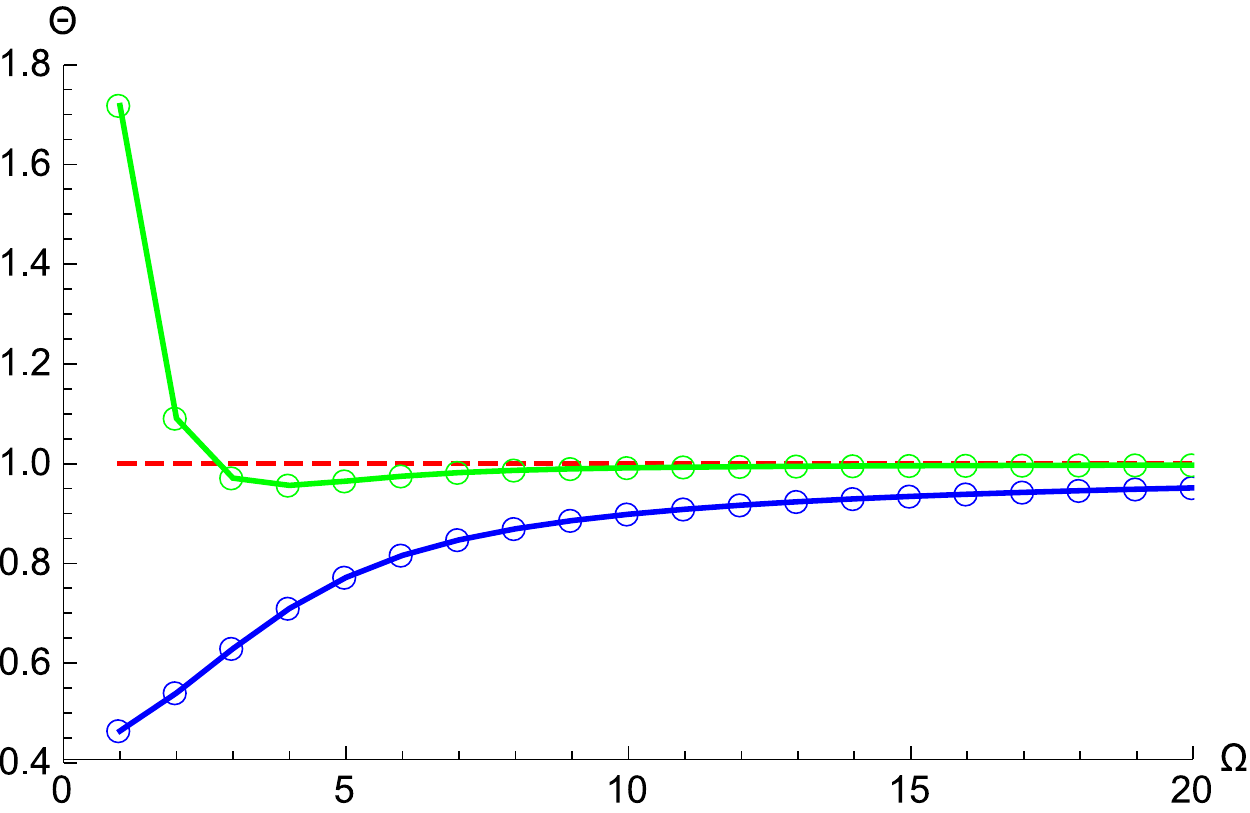}
\label{fig:subfig2}
}
\caption{Plots showing the dependence of the ratio of rate equation and CME mean molecule numbers ($\Delta$) and of the ratio of the LNA and CME Fano factors ($\Theta$) as a function of the volume $\Omega$ for the closed hetero-dimerization reaction $X_1 + X_2 \xrightleftharpoons[]{} X_3$ (blue, green) and for the open reaction $X_1 + X_2 \xrightleftharpoons[]{} X_3, \O  \xrightleftharpoons[]{} X_1$ (red dashed). See text for parameter values and for the method of calculation. Blue and green lines and open circles denote calculations for species $X_1$ and $X_3$, respectively; the red dashed line denotes calculations for both $X_1$ and $X_3$. Note that lines are simply a guide to the eye. Since $\Delta$ and $\Theta$ are generally not equal to one, for the closed reaction, it follows that the rate equations and the LNA differ from the CME's prediction of the first two moments for this reaction. In contrast $\Delta$ and $\Theta$ equal one for the open reaction, implying the LNA is exact for all volumes in this case.}
 \end{figure}

\section{A special class of systems}

\subsection{Specification}

Consider the following chemical system involving $N$ species interacting via $R$ reactions:
\begin{align}
\label{spsys}
\O  &\xrightleftharpoons[k_1^-]{k_1^+} X_1, \notag \\
\sum_{i=1}^N s_{ij} X_i &\xrightleftharpoons[k_j^-]{k_j^+} \sum_{i=1}^N r_{ij} X_i, \quad j = 2, ..., R,
\end{align}
where $k_j^{+/-} > 0$ and the following four constraints apply:
\begin{enumerate}
\item The Markov process describing the stochastic dynamics of the system is in detailed balance.
\item The stoichiometric integers $s_{ij}$ and $r_{ij}$ cannot simultaneously satisfy the two conditions $r_{1j} - s_{1j} = \pm 1$ and $r_{ij} - s_{ij} = 0$ for $j = 2, ...R$ and $i = 2, ..., N$.
\item The stoichiometric integers satisfy $1 \le \sum_i s_{ij} \le 2, 1 \le \sum_i r_{ij} \le 2, \ j = 2, ...R$.
\item Every second-order reaction must involve $X_1$.
\end{enumerate}

The first constraint implies that the frequency of transitions from a state $\vec{n}$ to another state $\vec{n}'$ equals the frequency of transitions from state $\vec{n}'$ to state $\vec{n}$, where $\vec{n} = \{n_1 , n_2 , ..., n_N \}$ and $n_i$ is the number of molecules of species $X_i$; note that this is not the same imposing the detailed balance condition on every pair of reversible reactions \cite{vanKampen}. The second constraint implies that reactions $j=2,..., R$ cannot lead to the production or destruction of one molecule of $X_1$ and simultaneously cause no change in the molecule numbers of other species. The latter restriction implies that reactions $j=2,..., R$ cannot for example be of the type $m X_1 \xrightleftharpoons[]{} (m+1) X_1$, $m \ge 1$. The third constraint implies the reactions in the system are first-order or second-order; this is since reactions involving three or more molecules are typically much rarer (at normal pressures and temperatures) than those involving one or two molecules. The fourth constraint requires that $X_1$ must participate in every second-order reaction in the system, either through the binding of two molecules of $X_1$ or else through the binding of one molecule of $X_1$ and of another species. 

Note that the detailed balance condition is independent of the other three constraints (see later for a further discussion of this point). Hence the four constraints taken together imply a subset of detailed balance systems. Our claim is that for the above constrained chemical system, the LNA exactly agrees with the CME up to second-order moments.

\subsection{Statistics of molecule number fluctuations of species $X_1$ according to the CME}

We start by showing that the {\emph{steady-state fluctuations of $X_1$ are Poissonian and uncorrelated with the fluctuations of all other species, including those which participate with $X_1$ in reversible unimolecular or second-order reactions}}. 

It is well known that the stochastic dynamics of the chemical system (\ref{spsys}) in well-mixed conditions is given by the CME \cite{vanKampen}:
\begin{align}
\label{cme}
  \partial_t P(\vec{n},t) 
  & = 
    \Omega k_1^+ P(n_1-1,n_2,..,n_N) - \Omega k_1^+ P(\vec{n}) \notag \\
   & + k_1^- (n_1 + 1) P(n_1+1,n_2,..,n_N) - k_1^- n_1 P(\vec{n}) \notag \\
    & + \Omega \sum_{j=2}^{R} \hat{f}_j^+ (\vec{n} - \vec{S}_j) P(\vec{n} - \vec{S}_j, t) - \Omega \sum_{j=2}^{R} \hat{f}_j^+ (\vec{n}) P(\vec{n}, t) \notag \\ & + 
    \Omega \sum_{j=2}^{R} \hat{f}_j^- (\vec{n} - \vec{S}_j) P(\vec{n} - \vec{S}_j, t) - \Omega \sum_{j=2}^{R} \hat{f}_j^- (\vec{n}) P(\vec{n}, t),
\end{align}
where $S_j$ is the jth column vector of the stoichiometric matrix $S$ (with elements $S_{ij} = r_{ij} - s_{ij}$) and $\Omega$ is the volume of the compartment in which the chemical system is confined. $P(\vec{n}, t)$ is the probability that at time $t$ the state of the system is described by the vector $\vec{n}$. From the law of mass action it follows that the (normalised) propensity functions $\hat{f}_j^{+/-}(\vec{n})$ of reaction $j$ (where $j = 2, ..., R $) are given by \cite{vanKampen}:
\begin{equation}
\label{props}
\hat{f}_j^+(\vec{n}) = k_j^+ \prod_{k=1}^N \frac{n_k!}{(n_k - s_{kj})! \Omega^{s_{kj}}}, \ \hat{f}_j^-(\vec{n}) = k_j^- \prod_{k=1}^N \frac{n_k!}{(n_k - r_{kj})! \Omega^{r_{kj}}}.
\end{equation}

Now by constraint 2, the only reaction which transitions between states $(n_1 , n_2 , ..., n_N)$ and $(n_1 \pm 1, n_2 , ..., n_N)$ is $\O  \xrightleftharpoons[]{} X_1$. Hence it follows by detailed balance (constraint 1) that in the CME we have the equivalence of two pairs of terms, namely:
\begin{align}
\label{Prelation}
 \Omega k_1^+ P (n_1 - 1, n_2 , ..., n_N ) &= k_1^- n_1 P (n_1 , n_2 , ..., n_N), \notag \\
 \Omega k_1^+ P (n_1, n_2 , ..., n_N ) &= k_1^- (n_1 + 1) P (n_1 + 1 , n_2 , ..., n_N).
\end{align} 
Hence the terms in Eq. (\ref{cme}) describing the reaction $\O  \xrightleftharpoons[]{} X_1$ sum to zero. The other terms describing reactions $j = 2, .., R$ sum independently to zero by the detailed balance condition since they describe transitions other than between the states $(n_1 , n_2 , ..., n_N)$ and $(n_1 \pm 1, n_2 , ..., n_N)$. Solving the recurrence relation given by Eq. (\ref{Prelation}), we obtain the marginal distribution $P(n_1) = \lambda^{n_1} e^{-\lambda} / n_1!$ where $\lambda = \Omega k_1^+/k_1^-$, i.e., the fluctuations in the molecule numbers of species $X_1$ are Poissonian with mean:
\begin{align}
\label{flucs1}
\langle n_1 \rangle &= \frac{\Omega k_1^+}{k_1^-}.
\end{align}
It also follows from Eq. (\ref{Prelation}) that 
\begin{align}
\label{flucs2}
\langle n_1 n_i \rangle &= \langle n_1 \rangle \langle n_i \rangle + \delta_{i,1} \langle n_1 \rangle, \quad i =1, ..., N, \\
\label{flucs3}
\langle n_1^2 n_i \rangle &= \langle n_1^2 \rangle \langle n_i \rangle + \delta_{i,1} \langle n_1 \rangle (1 + 2 \langle n_1 \rangle), \quad i =1, ..., N.
\end{align} 
Thus the fluctuations of $X_1$ are Poissonian and uncorrelated with the fluctuations of all other species in the system. 

We emphasize that constraint 2 is crucial to leading to Poissonian fluctuations in $X_1$. If for example we break this constraint, by considering the system of reactions $\O  \xrightleftharpoons[]{} X_1, X_1  \xrightleftharpoons[]{} 2 X_1$, then the detailed balance condition together with the CME implies that the fluctuations are non-Poissonian; these deviations from Poisson fluctuations are induced by the nonlinearity in the propensities of the reactions transitioning between states $(n_1 , n_2 , ..., n_N)$ and $(n_1 \pm 1, n_2 , ..., n_N)$. However it is to be emphasised that these non-Poissonian fluctuations are also uncorrelated with the fluctuations of all other species in the system. Hence to summarise, constraint 2 leads to Poissonian and uncorrelated fluctuations in $X_1$ while its lack leads to simply uncorrelated fluctuations in $X_1$.
 
\subsection{Statistics of molecule number fluctuations of all other species according to the CME} 
 
Next we use the CME to obtain equations for the mean molecule numbers and the fluctuations about these means for all species besides $X_1$.

By constraints 3 and 4, the reactions $j = 2, ..., R$ in our system are first or second-order and involve $X_1$ if they are second-order. This implies a simplification in the form of the propensities. To be specific, following are the three types of allowed chemical reactions and their associated propensities (for the jth reaction) as deduced using Eq. (\ref{props}): (i) a first-order unimolecular reaction involving the decay of some species $X_h$ is described by $\hat{f}_j^{+/-}(\vec{n}) = k_j^{+/-} n_h \Omega^{-1}$ where $h = 1, ..., N$; (ii) a second-order reaction between two molecules of $X_1$ is described by $\hat{f}_j^{+/-}(\vec{n})= k_j^{+/-} n_1 (n_1 - 1) \Omega^{-2}$ ; (iii) a second-order reaction between a molecule of $X_1$ and one of a different species $X_h$ is described by $\hat{f}_j^{+/-}(\vec{n}) = k_j^{+/-} n_1 n_h \Omega^{-2}$ . These three types of reactions can be generically captured by the propensity:
\begin{equation}
\label{gen_prop}
\hat{f}_j^{+/-} (\vec{n}) = k_j^{+/-} \biggl(\sum_{w=1}^N \alpha_{wj}^{+/-} \frac{n_w}{\Omega} + \sum_{w=2}^N \beta_{wj}^{+/-} \frac{n_1 n_w}{\Omega^2} + \gamma_{j}^{+/-} \frac{n_1 (n_1 - 1)}{\Omega^2} \biggr),
\end{equation}
where the $\alpha_{wj}^+$ equals one if the forward reaction $j$ is a unimolecular decay of a species $X_w$ and is otherwise zero, $\beta_{wj}^+$ equals one if the forward reaction $j$ is a second-order reaction between two different species $X_1, X_w$ and is otherwise zero and $\gamma_{j}^+$ equals one if the forward reaction $j$ is a second-order reaction between the same species $X_1$ and is otherwise zero. Similarly follows for the same coefficients but with minus superscript if the reverse reactions are of the type described.  

The time evolution of the mean concentration of species $X_i$, $\langle n_i \rangle /\Omega$, is obtained by multiplying the CME, Eq. (\ref{cme}), by $n_i/\Omega$ and summing over all $n_i$, leading to:
\begin{align}
\partial_t \frac{\langle n_i \rangle}{\Omega} & = \sum_{j=2}^R S_{ij} (\langle \hat{f}_j^+ (\vec{n}) \rangle -\langle \hat{f}_j^- (\vec{n}) \rangle) = 0, \\
&= \sum_{j=2}^R S_{ij} k_j^{+} \biggl(\sum_{w=1}^N \alpha_{wj}^{+} \frac{\langle n_w \rangle}{\Omega} + \sum_{w=2}^N \beta_{wj}^{+} \frac{\langle n_1 n_w \rangle}{\Omega^2} + \gamma_{j}^{+} \frac{\langle n_1 (n_1 - 1) \rangle}{\Omega^2} \biggr) \notag \\&- \sum_{j=2}^R S_{ij} k_j^{-} \biggl(\sum_{w=1}^N \alpha_{wj}^{-} \frac{\langle n_w \rangle}{\Omega} + \sum_{w=2}^N \beta_{wj}^{-} \frac{\langle n_1 n_w \rangle}{\Omega^2} + \gamma_{j}^{-} \frac{\langle n_1 (n_1 - 1) \rangle}{\Omega^2}\biggr), \label{means}
\end{align}
where the angled brackets denote the statistical average. Note that in the last line we substituted for $\hat{f}_j(\vec{n})$ from Eq. (\ref{gen_prop}). There is no dependence on $k_1^+$ and $k_1^-$ since as previously shown, the terms in the CME describing reaction $\O  \xrightleftharpoons[]{} X_1$ sum to zero. The time derivative is set to zero since detailed balance (constraint 1) implies steady-state conditions. Using the fact that fluctuations in $X_1$ are Poissonian and uncorrelated with the fluctuations of all other species, i.e., applying Eqs. (\ref{flucs1})-(\ref{flucs2}), we find that Eq. (\ref{means}) reduces to:
\begin{align}
\label{eq7}
\frac{d \phi_i}{dt} = 0 &= \sum_{j=2}^R S_{ij} k_j^+ \biggl(\sum_{w=1}^N \alpha_{wj}^+ \phi_w + \sum_{w = 2}^N \beta_{wj}^+ \phi_1 \phi_w+ \gamma_{j}^+ \phi_1^2 \biggr) \notag \\ & - \sum_{j=2}^R S_{ij} k_j^- \biggl(\sum_{w=1}^N \alpha_{wj}^- \phi_w + \sum_{w = 2}^N \beta_{wj}^- \phi_1 \phi_w+ \gamma_{j}^- \phi_1^2 \biggr),
\end{align}
where we denoted the concentration of species $X_i$, $\langle n_i \rangle / \Omega$, by $\phi_i$.

Next we derive equations for the second moments of the fluctuations about the mean concentrations. Let $C_{ik} = \Omega^{-2} (\langle n_i n_k \rangle - \langle n_i \rangle \langle n_k \rangle)$ be the covariance of the concentrations
fluctuations in species $X_i$ and $X_k$. It can be straightforwardly shown using the latter definition and Eq. (\ref{cme}) that the time-evolution of the covariance is given by:
\begin{align}
\label{cov_te}
\frac{d}{dt} C_{ik} &= 0 = \sum_{j=2}^R \biggl( \frac{S_{ij} S_{kj} \langle \hat{f}_j^+ (\vec{n}) + \hat{f}_j^- (\vec{n}) \rangle}{\Omega} + \frac{S_{kj}}{\Omega} (\langle n_i \hat{f}_j^+(\vec{n}) \rangle - \langle n_i \rangle \langle \hat{f}_j^+(\vec{n}) \rangle) + \frac{S_{ij}}{\Omega} (\langle n_k \hat{f}_j^+(\vec{n}) \rangle \notag \\ & - \langle n_k \rangle \langle \hat{f}_j^+(\vec{n}) \rangle) - \frac{S_{kj}}{\Omega} (\langle n_i \hat{f}_j^-(\vec{n}) \rangle - \langle n_i \rangle \langle \hat{f}_j^-(\vec{n}) \rangle) - \frac{S_{ij}}{\Omega} (\langle n_k \hat{f}_j^-(\vec{n}) \rangle - \langle n_k \rangle \langle \hat{f}_j^-(\vec{n}) \rangle) \biggr).
\end{align}

As before, this expression simplifies by the nature of the fluctuations in the number of molecules of species $X_1$. The first sum in Eq. (\ref{cov_te}) can hence be written as:
\begin{align}
\label{eq9}
D_{ik} (\vec{\phi}) = \sum_{j=2}^R S_{ij} S_{kj} \langle \hat{f}_j^+(\vec{n}) + \hat{f}_j^-(\vec{n}) \rangle &=  \sum_{j=2}^R S_{ij} S_{kj} k_j^+ \biggl(\sum_{w=1}^N \alpha_{wj}^+ \phi_w + \sum_{w = 2}^N \beta_{wj}^+ \phi_1 \phi_w+ \gamma_{j}^+ \phi_1^2 \biggr) \notag \\ & + \sum_{j=2}^R S_{ij} S_{kj} k_j^- \biggl(\sum_{w=1}^N \alpha_{wj}^- \phi_w + \sum_{w = 2}^N \beta_{wj}^- \phi_1 \phi_w+ \gamma_{j}^- \phi_1^2 \biggr).
\end{align}
The main term in the second sum in Eq. (\ref{cov_te}) can be written as:
\begin{align}
\langle n_i \hat{f}_j^+ (\vec{n}) \rangle - \langle n_i \rangle \langle \hat{f}_j^+(\vec{n}) \rangle &= k_j^+ \biggl( \sum_{w=1}^N \alpha_{wj}^+ \biggl( \frac{\langle n_i n_w \rangle}{\Omega} - \frac{\langle n_i \rangle \langle n_w \rangle}{\Omega} \biggr) + \sum_{w=2}^N \beta_{wj}^+ \biggl( \frac{\langle n_1 n_i n_w \rangle}{\Omega^2}  \notag \\ \label {eq10}&-\frac{\langle n_1 n_w \rangle \langle n_i \rangle}{\Omega^2} \biggr) + \gamma_{j}^+ \biggl( \frac{\langle n_i n_1 (n_1 - 1) \rangle}{\Omega^2} - \frac{\langle n_1 (n_1 - 1) \rangle \langle n_i \rangle}{\Omega^2} \biggr) \biggr)\\ \label{eq11} &= k_j^+ \Omega (\sum_{w=1}^N \alpha_{wj}^+ C_{iw} + \phi_1 \sum_{w=2}^N \beta_{wj}^+ C_{iw} + C_{i1} \sum_{w=2}^N \beta_{wj}^+ \phi_w + 2 \gamma_{j}^+ \phi_1 C_{i1}),
\end{align}
where Eq. (\ref{eq11}) follows from Eq. (\ref{eq10}) by the application of Eqs. (\ref{flucs1})-(\ref{flucs3}). The first sum and the last term in Eq. (\ref{eq11}) can be easily derived. The second sum in Eq. (\ref{eq11}) is less straightforward to obtain and hence we provide some additional intermediate steps, as follows. One first considers the third cumulant $K_{1wi}$ which by definition is: 
\begin{align}
\label{eq12a}
K_{1wi} = \langle n_1 n_w n_i \rangle - \langle n_w n_i \rangle \langle n_1 \rangle - \langle n_i \rangle \langle n_1 n_w \rangle - \langle n_w \rangle \langle n_1 n_i \rangle + 2 \langle n_1 \rangle \langle n_w \rangle \langle n_i \rangle.
\end{align}
Since $w \ne 1$ (as the second sum in Eq. (\ref{eq10}) is from 2 to $N$), it follows that the possible cases are $i = 1, w \ne 1$ and $i \ne 1, w \ne 1$. It is easy to verify using Eqs. (\ref{flucs2})-(\ref{flucs3}) that for each of these two cases, $K_{1wi} = 0$. Hence we have:
\begin{align}
\label{eq12}
\langle n_1 n_i n_w \rangle - \langle n_1 n_w \rangle \langle n_i \rangle &= \langle n_1 \rangle ( \langle n_w n_i \rangle - \langle n_i \rangle \langle n_w \rangle) + \langle n_w \rangle(\langle n_i n_1 \rangle - \langle n_1 \rangle \langle n_i \rangle) \notag \\ &= \Omega^3 (\phi_1 C_{iw} + \phi_w C_{i1}),
\end{align}
which leads to the second and third sums in Eq. (\ref{eq11}). 

Hence using Eq. (\ref{eq9}) and Eq. (\ref{eq11}), we can deduce that the equation for the covariance of fluctuations, Eq. (\ref{cov_te}), reduces to:
\begin{align}
\label{eq14}
\frac{d}{dt} C_{ik} = 0 = \frac{D_{ik} (\vec{\phi})}{\Omega} + \sum_{j=2}^R S_{kj} (\Lambda_{ij}^+ - \Lambda_{ij}^-) + \sum_{j=2}^R S_{ij} (\Lambda_{kj}^+ - \Lambda_{kj}^-),
\end{align}
where $\Lambda_{ij}^+ \Omega$ equals Eq. (\ref{eq11}) and $\Lambda_{ij}^- \Omega$ denotes the same but with minus superscripts. 


The importance of the four constraints is now clear from the derivation in this subsection and the previous. Constraint 1 (detailed balance) leads to fluctuations in the molecule number of species $X_1$ to be uncorrelated with the fluctuations in the molecule numbers of all other species. Constraint 2 leads to Poissonian fluctuations in the molecule number of species $X_1$. Constraint 3 leads to a simple quadratic form for the propensities which considerably simplifies the calculation. Constraint 4 leads to the equations for the second moments in all species to only depend on those third-order cumulants which involve $X_1$. By constraints 1, 2 and 3 we have shown that the equations for the mean molecule numbers of all species, Eqs. (\ref{eq7}), are uncoupled from the second and higher-order moments; this is since uncorrelated fluctuations forces the condition $\langle n_1 n_w \rangle = \langle n_1 \rangle \langle n_w \rangle$ for $w \ne 1$ while Poissonian fluctuations forces the condition $\langle n_1 (n_1 - 1) \rangle = \langle n_1 \rangle^2$. By constraints 1 and 4 we have shown that the equations for the second-moments of the molecule numbers of all species, Eqs. (\ref{eq14}), are uncoupled from the third and higher-order moments; this is since all the third-order cumulants involve $X_1$ and hence must be zero since species $X_1$ is uncorrelated from the rest. Thus the importance of the four constraints is that together they lead to the equations for the moments to naturally decouple from the higher-order moments. As we shall see, this property is crucial for achieving agreement of the CME with the rate equations and the LNA, since the rate equations depend only on the concentrations (not on the second-moments) while the LNA equations for the second-moments depend only on the concentrations and on the second-moments (not on the third-moments).

Note that detailed balance by itself would not have led to the decoupled equations that we obtained, since detailed balance cannot generally guarantee Poissonian correlations in $X_1$ (or in any other species for that matter -- see for example \cite{vanKampen1976}), and since generally the second-moment equations will depend on third-order cumulants which do not all involve $X_1$ (see for example Appendix C of \cite{GrimaJCP2012}).

An important point worth mentioning regarding our derivation is that we did not use information about the third and higher-order moments of $X_1$ and hence the {\emph{derivation holds also if the fluctuations in $X_1$ were not Poissonian but only agreed with a Poissonian up to second-order moments}}.

\subsection{Statistics of molecule number fluctuations using the LNA}

Next we derive equations for the same quantities using the LNA and show that they are one and the same as the equations we just obtained using the CME. The LNA has been extensively discussed previously (see for example \cite{vanKampen,ElfEhrenberg2003}) and thus here we shall simply state the main results. 

As we saw earlier, constraints 1 and 2 imply that the reaction $\O  \xrightleftharpoons[]{} X_1$ can be treated as if it is separate and non-interacting with the rest of the reactions ($j = 2, ..., R$) in chemical system (\ref{spsys}). In particular the means and fluctuations of $X_1$ can be found by considering the CME of reaction $\O  \xrightleftharpoons[]{} X_1$ while the means and fluctuations of the rest of the species are found by considering the CME of reactions $\sum_{i=1}^N s_{ij} X_i \xrightleftharpoons[]{} \sum_{i=1}^N r_{ij} X_i$ where $j = 2, ..., R$. Since the LNA is an approximation of the CME, it follows that we can apply the LNA to each of the two CMEs. 

Now it is well known that the LNA agrees with the CME up to second-order moments for systems of at most first-order reactions. Hence the application of the LNA to the CME of $\O  \xrightleftharpoons[]{} X_1$ leads to the same mean number of molecules and second-moments of the fluctuations about this mean for species $X_1$ as found earlier using the CME approach (see Eqs.(\ref{flucs1})-(\ref{flucs2})). 

Next we apply the LNA to the CME of reactions $j = 2, ..., R$ to obtain the means and fluctuations of the concentrations of all other species besides $X_1$. Given the chemical reaction system (\ref{spsys}) and assuming that the system is deterministically monostable (see later for a simple proof of this property), the LNA states that the time-evolution of the mean concentrations is given by the conventional rate equations \cite{vanKampen}:
\begin{equation}
\label{eq15}
\frac{d}{dt} \phi_i = \sum_{j=2}^R S_{ij} (f_j^+ (\vec{\phi}) - f_j^- (\vec{\phi})) = 0,
\end{equation}
where $f_j^{+/-}(\vec{\phi})$ is the macroscopic rate vector for the forward ($+$) or the backward ($-$) reaction $j$ as given by the law of mass action. Setting the time derivative to zero follows by the fact that constraint 1 implies steady-state conditions. Given constraints 3 and 4, the reactions $j = 2, ..., R$ in our system are first or second-order and involve $X_1$ if they are second-order. This implies, by the law of mass action, that the macroscopic rate vector takes the following form: (i) a first-order unimolecular reaction involving the decay of some species $X_h$ is described by $\hat{f}_j^{+/-}(\vec{\phi}) = k_j^{+/-} \phi_h$ where $h = 1, ..., N$; (ii) a second-order reaction between two molecules of $X_1$ is described by $\hat{f}_j^{+/-}(\vec{\phi})= k_j^{+/-} \phi_1^2$ ; (iii) a second-order reaction between a molecule of $X_1$ and one of a different species $X_h$ is described by $\hat{f}_j^{+/-}(\vec{
\phi}) = k_j^{+/-} \phi_1 \phi_h$. These three types of reactions can be generically captured by the rate vector:
\begin{align}
\label{eq16}
f_j^{+/-} (\vec{\phi}) = k_j^{+/-} \biggl(\sum_{m=1}^N \alpha_{mj}^{+/-} \phi_m + \sum_{w = 2}^N \beta_{wj}^{+/-} \phi_1 \phi_w+  \gamma_{j}^{+/-} \phi_1^2 \biggr),
\end{align}
where the $\alpha$'s, $\beta$'s and $\gamma$'s are either 0 or 1, depending which one of the three elementary
interactions above describes reaction $j$. Note that Eq. (\ref{eq15}) together with Eq. (\ref{eq16}) leads to the final equations determining the mean concentrations according to the LNA (and rate equations) and these are precisely the same as those previously obtained from the CME (see Eq. (\ref{eq7})). 

As previously mentioned the LNA is only applicable if the rate equations are monostable. Now as we saw earlier, by constraints 1 and 2 there is only one steady-state value for the mean molecule number of species $X_1$. Furthermore Eq. (\ref{eq15}) together with Eq. (\ref{eq16}) are linear in the concentrations $\phi_i$ ($i = 2, .., N$) which implies one steady-state solution for the concentrations of all species (monostability). 

Under the LNA, the covariance of concentration fluctuations is described by the Lyapunov equation \cite{ElfEhrenberg2003}:
\begin{align}
\label{eq17}
\frac{d}{dt} C_{ik} = 0 = \frac{\sum_{j=2}^R S_{ij} S_{kj} (f_j^+ (\vec{\phi})+f_j^- (\vec{\phi}))}{\Omega} + \sum_{w=1}^N (J_{iw} C_{wk} + J_{kw} C_{iw}), 
\end{align}
where $J_{kw} = \sum_{j=2}^R S_{kj} \frac{d}{d \phi_w} (f_j^+ (\vec{\phi})-f_j^- (\vec{\phi})) = J_{kw}^+ - J_{kw}^-$ is the $(k, w)$ element of the Jacobian matrix of the rate equations given by Eq. (\ref{eq15}). Using Eq. (\ref{eq16}) and the definition of $J_{kw}^{+/-}$ above, we find
\begin{align}
\sum_{w=1}^N J_{kw}^{+/-} C_{iw} &= \sum_{w=1}^N \sum_{j=2}^R k_j^{+/-} S_{kj} \bigl( \alpha_{wj}^{+/-} + \delta_{w,1} \sum_{s=2}^N \beta_{sj}^{+/-} \phi_s + (1 - \delta_{w,1}) \beta_{wj}^{+/-} \phi_1 + 2 \gamma_{j}^{+/-} \phi_1  \delta_{w,1}  \bigr) C_{iw}, \notag \\ \label{eq19} &= \sum_{j=2}^R S_{kj} \Lambda_{ij}^{+/-},
\end{align}
where $\Lambda_{ij}^{+/-}$ is as defined earlier after Eq. (\ref{eq14}). Note that here, use was made of the relation $C_{i1} = C_{11} \delta_{i,1}$, which is a statement of the properties of the fluctuations of $X_1$. 

Given Eq. (\ref{eq16}), it is easy to verify that the first term in Eq. (\ref{eq17}) is given by:
\begin{align}
\label{eq18a}
D_{ik} (\vec{\phi}) = \sum_{j=2}^R S_{ij} S_{kj} (f_j^+ (\vec{\phi})+f_j^- (\vec{\phi})),
\end{align}
where $D_{ik} (\vec{\phi})$ is as defined in Eq. (\ref{eq9}).

Finally substituting  Eq. (\ref{eq19}) and Eq. (\ref{eq18a}) in Eq. (\ref{eq17}) we obtain Eq. (\ref{eq14}), i.e., the equations for the covariance of concentration fluctuations according to the LNA are one and the same as the equations obtained earlier using the CME.  

Note that the third and higher-order moments of the concentration fluctuations according to the LNA do not agree with those of the CME. This is since the LNA provides a Gaussian approximation to the probability distribution solution of the CME \cite{vanKampen}, i.e., it leads to third and higher-order cumulants equal to zero, whereas the Poissonian fluctuations of the species interacting in second-order reactions are characterised by non-zero cumulants to all orders.

\section{Generalisation and some examples}

Our proof in the previous section has been for the system of reactions (\ref{spsys}) in which species $X_1$ is special, in the sense that it is being produced and destroyed by a particular reaction, and participates in all second-order reactions. Relaxing this speciality of $X_1$ leads to a broader class of systems for which the LNA is exact up to second-order moments. 

In particular consider the following chemical system of $N$ species interacting via $R$ reactions:
\begin{align}
\label{spsys1}
&\O  \xrightleftharpoons[k_1^-]{k_1^+} X_{i_1}, \O  \xrightleftharpoons[k_2^-]{k_2^+} X_{i_2}, ...., \O  \xrightleftharpoons[k_S^-]{k_S^+} X_{i_S}, \notag \\
&\sum_{i=1}^N s_{ij} X_i \xrightleftharpoons[k_j^-]{k_j^+} \sum_{i=1}^N r_{ij} X_i, \quad j = S+1, ..., R,
\end{align}
where $k_j^{+/-} > 0$, $1 \le S \le R - 1$, and $i_1, ..., i_S$ are positive integers taking a value between $1$ and $N$. We also apply the following constraints: 
\begin{enumerate}
\item The Markov process describing the stochastic dynamics of the system is in detailed balance.
\item The stoichiometric integers $s_{ij}$ and $r_{ij}$ cannot simultaneously satisfy the two conditions $r_{i_k,j} - s_{i_k,j} = \pm 1$ and $r_{i_k,j} - s_{i_k,j} = 0$ for $j = S+1, ...R$ and $k = 1, ..., S$.
\item The stoichiometric integers satisfy $1 \le \sum_i s_{ij} \le 2, 1 \le \sum_i r_{ij} \le 2, \ j = 2, ...R$.
\item Every second-order reaction involves at least one of the following species $X_{i_1}$, $X_{i_2}$, ..., $X_{i_S}$.
\end{enumerate}

This system is a generalisation of the system (\ref{spsys}) since the properties previously particular to species $X_1$ are now common to $S$ different species, $X_{i_1}$, $X_{i_2}$, ..., $X_{i_S}$. Constraint 2 implies that the reactions $\O  \xrightleftharpoons[]{} X_{i_1}, \O  \xrightleftharpoons[]{} X_{i_2}, ...., \O  \xrightleftharpoons[]{} X_{i_S}$ are the only reactions in the system which lead to the increase or decrease of one molecule of the species $X_{i_1}$, $X_{i_2}$, ..., $X_{i_S}$ while causing no change to the number of molecules of all other species. 

The proof that the LNA is exact up to second-order moments for system (\ref{spsys1}) follows along the lines of the proof provided in the previous section. Constraints 1 and 2 lead to steady-state fluctuations in the concentrations of species $X_{i_1}$, $X_{i_2}$, ..., $X_{i_S}$ which are Poissonian and uncorrelated with the fluctuations in the concentrations of all other species; these properties together with the special form of the propensities due to constraints 3 and 4 lead to the exactness of the LNA in detailed balance conditions. 

A perusal of the proof provided in the previous section shows that the crucial ingredient for the exactness of the LNA up to second-order moments for any chemical system with up to second-order reactions is that {\emph{the fluctuations in at least one of the species participating in each second-order reaction are Poissonian (up to second-order moments) and uncorrelated with the fluctuations of all other species}}; the constrained chemical systems (\ref{spsys}) and (\ref{spsys1}) are examples of such systems but its unlikely that they are the only ones satisfying the aforementioned crucial ingredient. Furthermore it can also be deduced that {\emph{if the chemical system only has second-order reactions between different species, i.e, $\gamma_j^{+/-} = 0$ in Eqs. (\ref{gen_prop}) and (\ref{eq16}), then exactness of the LNA is guaranteed if the fluctuations in at least one of the species participating in each second-order reaction are uncorrelated with the fluctuations of all other species (no Poissonian fluctuations restriction).}}

A list of exemplary chemical systems of the type (\ref{spsys1}) and satisfying the above four constraints, is as follows:
\begin{align}
\label{eq21}
\O  &\xrightleftharpoons[k_1]{k_0} X_1, X_1 + X_2 \xrightleftharpoons[k_3]{k_2} X_3, \\
\label{eq22}
\O  &\xrightleftharpoons[k_1]{k_0} X_1, X_1 + X_1 \xrightleftharpoons[k_3]{k_2} X_2, \\
\label{eq23}
\O  &\xrightleftharpoons[k_1]{k_0} X_1, X_1 + X_2 \xrightleftharpoons[k_3]{k_2} 2 X_1, \\
\label{eq24}
\O  &\xrightleftharpoons[k_1]{k_0} X_1, \O  \xrightleftharpoons[k_3]{k_2} X_2, X_1 + X_2 \xrightleftharpoons[k_5]{k_4} 2 X_2, \\
\label{eq25}
\O  &\xrightleftharpoons[k_1]{k_0} X_2, X_1 + X_2 \xrightleftharpoons[k_3]{k_2} X_3 \xrightleftharpoons[k_5]{k_4} X_2 + X_4 , \\
\label{eq26}
\O  &\xrightleftharpoons[k_1]{k_0} X_1, X_1 + X_i \xrightleftharpoons[k_{2i+1}]{k_{2i}} X_{i+1}, \quad i = 1,..., N - 1.
\end{align}
These reactions describe heterodimerisation (\ref{eq21}) and homodimerisation (\ref{eq22}), autocatalytic reactions (\ref{eq23})-(\ref{eq24}), an enzyme reaction (\ref{eq25}) and a polymerisation reaction (\ref{eq26}) leading to a polymer made of $N$ monomers. 

In Appendix A, as a secondary check, we explicitly solve the steady-state CME of the six reactions (\ref{eq21}-\ref{eq26}) in detailed balance conditions. It can be straightforwardly verified using these steady-state distributions that the mean and variance of the molecule number fluctuations of all species exactly agree with those obtained from the rate equations and the LNA, respectively. From the exact solutions once can also verify that (i) the fluctuations  of {{\emph{all species}} in the system are Poissonian and uncorrelated for chemical systems (\ref{eq22}-\ref{eq24}) and (\ref{eq26}); this is since in each of these cases there are no implicit chemical conservation laws. (ii) only the fluctuations of {{\emph{one of the species involved in each second-order reaction is Poissonian and uncorrelated}} for systems (\ref{eq21}) and (\ref{eq25}); this is since these systems do have implicit chemical conservation laws. 

It has been shown in \cite{Anderson2010} that the network property called deficiency has implications for the form of the steady-state distribution solution of the CME and hence one might ponder a possible link between deficiency and the exactness of the LNA.  In particular Anderson et al. showed that systems which are weakly reversible, have a deficiency of zero and lack any implicit conservation laws are in detailed balance and characterised by fluctuations in the molecule numbers of all species which are Poissonian and uncorrelated, and hence by the results of Section III it follows that for such systems, the LNA is exact up to second-order moments. Reactions satisfying such criteria are (\ref{eq22}), (\ref{eq23}) and (\ref{eq26}). However generally systems of the type (\ref{spsys1}) and with the four constraints delineated above do not possess a deficiency of zero and they do have implicit conservation laws. For example reaction (\ref{eq24}) has a deficiency of one while reactions (\ref{eq21}) and (\ref{eq25}) do have implicit conservation laws.

One may also ponder whether the exact agreement up to second-order moments between the rate equations and the LNA and the CME for systems of type (\ref{spsys1}) in steady-state and detailed-balance conditions also extends for the whole time-evolution of the system, i.e., can we relax constraint 1? Simulations of the chemical reaction system (\ref{eq22}) using the stochastic simulation algorithm \cite{Gillespie1977} confirm that this is not the case (see Fig. 2), i.e., exact agreement is found only in steady-state conditions. As expected, the error made by the LNA for finite time decreases as the volume of the system increases.

\begin{figure}
\centering
\subfigure[]{
\includegraphics[width=92mm]{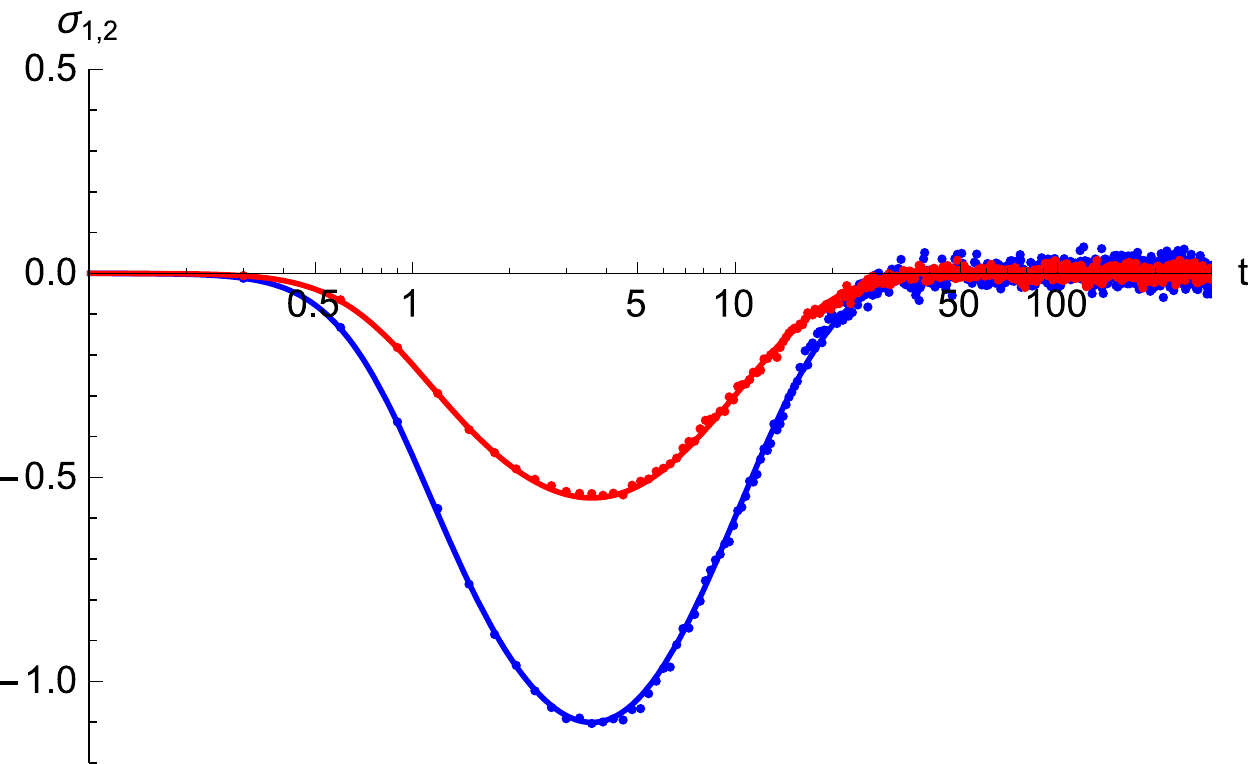}
\label{fig:subfig1}
}
\subfigure[]{
\includegraphics[width=92mm]{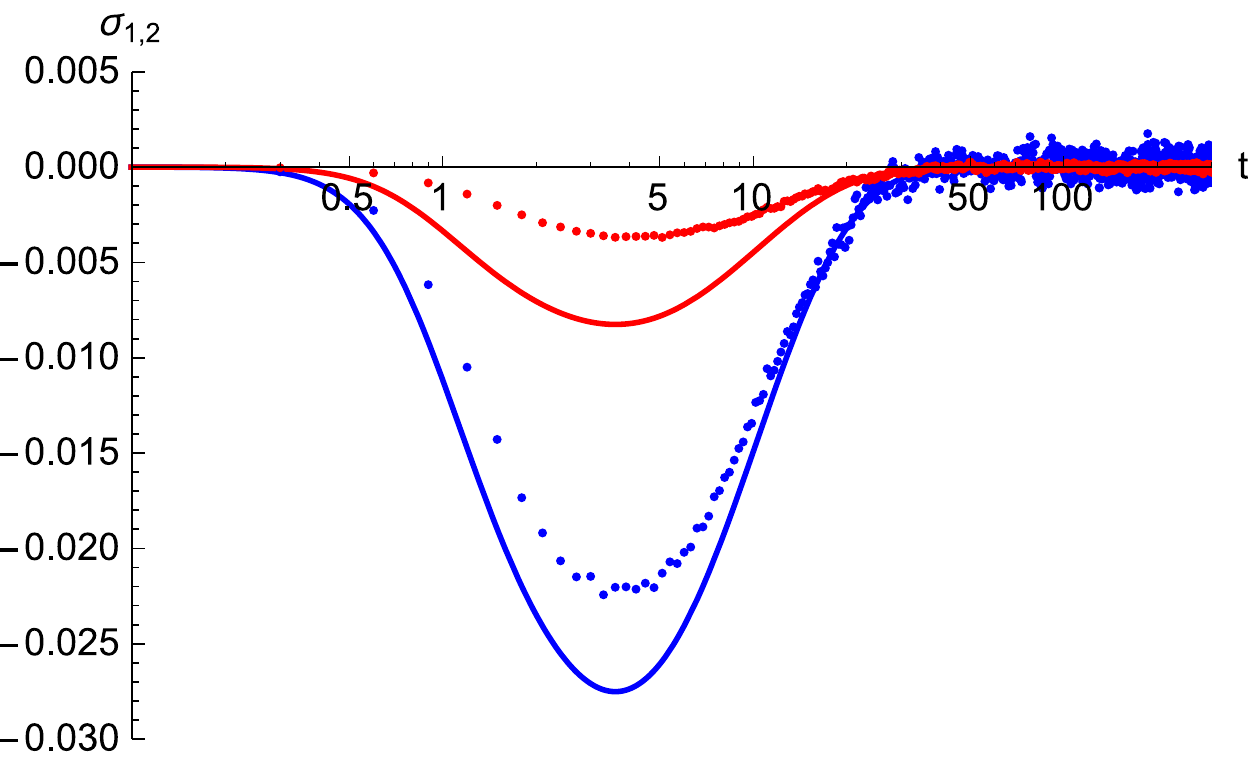}
\label{fig:subfig2}
}
\caption{Plots of the time dependence of the covariance $\sigma_{1,2}$ in the fluctuations of species $X_1$ and $X_2$ in reaction system (\ref{eq22}). Solid lines show the solution of the LNA. The dots show an ensemble average over a large number of trajectories generated using the stochastic simulation algorithm. The rate constants are $k_0 = k_1 = k_2 = k_3 = 1$. The volumes are $\Omega = 20$ (red) and $\Omega = 40$ (blue) in (a) and $\Omega = 0.3$ (red) and $\Omega = 1$ (blue) in (b). Note that the LNA agrees with the CME in steady-state conditions for all volumes; however the time-evolution predicted by the LNA is not in exact agreement with the CME and the discrepancy increases as the volume is decreased from $\Omega = 40$ to $0.3$.}\label{fig:plots}
 \end{figure}

\subsection{Spatially extended systems}}

Up till now we have specifically interpreted each $X_i$ as a unique chemical species, e.g. in the context of gene expression, $X_1$ might refer to the mRNA whereas $X_2$ might refer to the protein which is translated by the mRNA. Within this interpretation, the CME Eq. (\ref{cme}) and the corresponding LNA given by Eq. (\ref{eq17}) provide a non-spatial stochastic description of the dynamics of the well-mixed system of $N$ unique chemical species interacting via the reactions (\ref{spsys}). However by changing the manner in which one interprets $X_i$, one can also apply the results derived earlier to spatially extended systems.

Lets say we want to model a second-order reaction $A + B \xrightleftharpoons[]{} C$ between a species $A$ which diffuses and two non-diffusing species $B$ and $C$ which are localised in a certain part of space. This reaction could, for example, be occurring inside a cell and the localisation could be due to a membrane-bound compartment in which the membrane selectively allows through it only certain species, in this case species $A$ \cite{Albert}.

For simplicity lets assume that space is divided into 3 equally sized subvolumes (or voxels) and that $A$ is free to move between the three voxels while $B$ and $C$ are localised in one voxel. A scheme describing this spatially extended system is $\O \xrightleftharpoons[]{} X_1 \xrightleftharpoons[]{} X_2 \xrightleftharpoons[]{} X_3$, $X_1 + X_4 \xrightleftharpoons[]{} X_5$ where $X_1$, $X_2$ and $X_3$ denote the same chemical species (species $A$) in voxels $1$, $2$ and $3$ respectively, while $X_4$ and $X_5$ denote species $B$ and $C$ respectively. Specifically the aforementioned two sets of reactions respectively model the diffusion of species $A$ from the surrounding space into voxel 1 and its subsequent diffusion into voxels 2 and 3, and the chemical interaction of species $A$ with species $B$ and $C$ localised in voxel 1.

The stochastic dynamics of this system (assuming well-mixing in each voxel) is still described by Eq. (\ref{cme}) which is now referred to as the reaction-diffusion master equation (RDME), with the proviso that the volume $\Omega$ is now the volume of each voxel \cite{Gillespie2013}. The approximate stochastic dynamics of this system is also still described by the Lyapunov Eq. (\ref{eq17}) (sometimes referred to as the multi-compartment LNA \cite{Challenger2012}). Additionally this spatially-extended reaction system satisfies the four constraints mentioned in Section III.A and hence it follows by the results of Section III that the multi-compartment LNA and the RDME agree to second-order moments in the molecule number fluctuations in each of the boxes composing space. It is straightforward to show that this equivalence of the two modelling approaches also holds if: (i) space is divided into any number of voxels; (ii) one lets species $B$ and $C$ diffuse freely throughout space; (iii) the diffusive entry of any species, from the surroundings into the space under consideration, occurs at any (or all) of the voxels.

The reaction here considered is a spatially extended version of reaction (\ref{eq21}); similarly one can write spatially-extended versions of reactions (\ref{eq22})-(\ref{eq26}) and in each case one finds the equivalence of the multi-compartment LNA and the RDME up to second-order moments. By similar arguments to the one above, one can deduce that a spatially extended system involving $N$ chemical species of which $S$ of them diffuse from the surroundings into the space of interest, and diffuse within this space to participate in a number of reactions is a special case of system (\ref{spsys1}). If the chemical reactions within this space also satisfy the four constraints delineated previously, then the agreement of the multi-compartment LNA and the RDME up to second-order moments is guaranteed.

\section{Discussion \& Conclusion}

In this article, we have shown that for a class of chemical systems, the LNA gives results for the mean concentrations and second moments of fluctuations about the means which exactly match those given by the CME for all volumes, even though the chemical system is composed of at least one second-order reaction. In particular this also implies that the mean concentrations of the CME are in exact agreement with those of the conventional rate equations, on which the LNA is based. This is in contrast to the current prevalent thinking that posits the LNA is only exact for a chemical system with zero and first-order reactions. Our results are also in contrast to the fact that previous studies have found strong deviations between the rate equations and the CME \cite{Grima2010,Thomas2010} and between the LNA and the CME for various biochemical systems involving second-order reactions \cite{Thomas2013,Thomas2014}. For example in \cite{Thomas2010} it has been found that the differences between the substrate concentrations predicted by the rate equations and by the CME increase as one goes further downstream in certain large enzyme reaction systems and in \cite{Thomas2013} it was shown how the dependence of the coefficient of variation of protein noise versus the stress level in a gene regulatory circuit according to the LNA is roughly parabolic while the same obtained from the CME is a monotonic increasing function. What is clear from the results of this paper and the aforementioned results in the literature, is that the wiring of the chemical reaction network plays a major role in determining the deviations between the LNA and the CME, when there are second-order reactions. In other words, the differences between the predictions of the rate equations / LNA and of the CME are proportional to the elements of the Hessian matrix of the rate equations and also to a ``wiring factor''; hence the deviations from the LNA are zero either if the Hessian is zero, i.e, if the chemical system is composed of zero and first-order reactions (the well known case) or if the system has up to second-order reactions but the wiring factor is zero (the case elucidated in this paper). The four constraints on our system impose fairly stringent requirements on the wiring of the network and hence if they turn out to be necessary for the exactness of the LNA (remains to be proved) then this property is unlikely to be common to many realistic chemical and biochemical networks.

Our derivation in Section III clarifies that the crucial property of the chemical systems for which there is an equivalence of the LNA and the CME up to second-order moments is that for {\emph{at least one of the species participating in each second-order reaction, the steady-state fluctuations in the molecule numbers are Poissonian (up to second-order moments) and uncorrelated with the fluctuations in the molecule numbers of all species}}. Note that this property, does not exclude the possibility that one of the species participating in a second-order reaction (or other reaction) has non-Poissonian and correlated fluctuations. For example in reaction (\ref{eq21}), there is the implicit conservation law $n_2 +n_3$ = constant which leads to correlation in the fluctuations of the molecule numbers of $X_2$ and $X_3$, while in reaction (\ref{eq25}), we have the conservation law $n_1 +n_3 +n_4$ = constant which leads to correlated fluctuations in the molecule numbers of species $X_1, X_3, X_4$. We emphasise that for such cases, the exactness of the LNA extends to all species in the system, not just those exhibiting Poissonian and uncorrelated fluctuations.

We have shown that the aforementioned crucial property leads to agreement between the CME and LNA up to second-order moments because it leads to a truncation of the usually infinite hierarchy of coupled moment equations one obtains from the CME. This truncation is unlike that of moment-closure approximation methods \cite{Verghese2007,Ale2013}, in the sense that in the latter one imposes an ad-hoc assumption to artificially truncate the moment equations whereas in our case we have shown that the truncation follows directly and automatically from the properties of a subset of detailed balanced chemical systems and hence is exact. An interesting question remains as to whether the detailed balance condition can be relaxed or if it is a necessary condition to guarantee the exactness of the LNA.

\begin{appendix}

\section{Detailed balance solutions of the CME of chemical systems (\ref{eq21}-\ref{eq26})}
\begin{align}
\label{eqA1}
P(n_1,n_2) &= e^{-\frac{k_0 \Omega}{k_1}} \biggl(1 + \frac{k_1 k_3}{k_0 k_2} \biggr)^{-n_T} n_T! \frac{(\frac{k_0 \Omega}{k_1})^{n_1}}{n_1!} \frac{(\frac{k_1 k_3}{k_0 k_2})^{n_2}}{n_2! (n_T - n_2)!}, \\
\label{eqA2}
 P(n_1,n_2) &= e^{-\frac{k_0 \Omega}{k_1}} e^{-\frac{k_0^2 k_2 \Omega}{k_1^2 k_3}} \frac{(\frac{k_0 \Omega}{k_1})^{n_1}}{n_1!} \frac{(\frac{k_0^2 k_2 \Omega}{k_1^2 k_3})^{n_2}}{n_2!}, \\
 \label{eqA3}
P(n_1,n_2) &= e^{-\frac{k_0 \Omega}{k_1}} e^{-\frac{k_0 k_3 \Omega}{k_1 k_2}} \frac{(\frac{k_0 \Omega}{k_1})^{n_1}}{n_1!} \frac{(\frac{k_0 k_3 \Omega}{k_1 k_2})^{n_2}}{n_2!}, \\
\label{eqA4}
P(n_1,n_2) &= e^{-\frac{k_0 \Omega}{k_1}} e^{-\frac{k_2 \Omega}{k_3}} \frac{(\frac{k_0 \Omega}{k_1})^{n_1}}{n_1!} \frac{(\frac{k_2 \Omega}{k_3})^{n_2}}{n_2!}, \\
\label{eqA5}
P(n_1,n_2,n_4) &= e^{-\frac{k_0 \Omega}{k_1}}  \frac{(\frac{k_0 \Omega}{k_1})^{n_2}}{n_2!}  \biggl(\biggl(1 + \frac{k_1 k_3}{k_0 k_2} \biggr) \biggl(1 + \frac{k_1 k_2 k_4}{(k_0 k_2 + k_1 k_3) k_5} \biggr)\biggr)^{-n_T} \frac{n_T! (\frac{k_1 k_3}{k_0 k_2})^{n_1 + n_4} (\frac{k_2 k_4}{k_3 k_5})^{n_4}}{n_1! n_4! (n_T - n_1 - n_4)!}, \\
\label{eqA6}
P(n_1,...,n_N) &= e^{-\frac{k_0 \Omega}{k_1}} ... e^{({-\frac{k_0 \Omega}{k_1}})^N (\prod_{w=1}^{N-1} \frac{k_{2w}}{k_{2w+1}}) \Omega} \frac{(\frac{k_0 \Omega}{k_1})^{n_1}}{n_1!} ... \frac{((\frac{k_0 \Omega}{k_1})^{N} (\prod_{w=1}^{N-1} \frac{k_{2w}}{k_{2w+1}})\Omega)^{n_N}}{n_N!}.
\end{align}
The fluctuations are Poissonian for species $X_1$ in Eq. (\ref{eqA1}), for species $X_1$ and $X_2$ in Eq. (\ref{eqA2})-(\ref{eqA4}), for species $X_2$ in Eq. (\ref{eqA5}) and for species $X_1$ to $X_N$ in system (\ref{eqA6}). Note that the non-Poissonian marginal distributions of some species ($X_2$ in Eq. (\ref{eqA1}); $X_1$ and $X_4$ in Eq. (\ref{eqA5})) only originate when there is an implicit conservation law e.g. $n_2 +n_3 = n_T$ = constant in Eq. (\ref{eqA1}) and $n_1 +n_3 +n_4 = n_T$ = constant in Eq. (\ref{eqA5}). The exact solutions are obtained using the method expounded in \cite{vanKampen1976}.

\end{appendix}


\begin{thebibliography}{}
\bibitem{McQuarrie1967} D. A. McQuarrie, J. App. Prob. {\bf{4}}, 413 (1967)
\bibitem{Verghese2007} C. A. Gomez-Uribe and G. C. Verghese, J. Chem. Phys. {\bf{126}}, 024109 (2007)
\bibitem{Ale2013} A. Ale, P. Kirk and M. P. H. Stumpf, J. Chem. Phys. {\bf{138}}, 174101 (2013)
\bibitem{vanKampen} N. G. van Kampen, {\it Stochastic Processes in Physics and Chemistry} (Elsevier, 2007)
\bibitem{vanKampen1961} N. van Kampen, Can. J. Phys. {\bf{39}}, 551 (1961)
\bibitem{GrimaJCP2012} R. Grima, J. Chem. Phys. {\bf{136}}, 154105 (2012)
\bibitem{Ferm2008} L. Ferm, P. L\"otstedt and A. Hellander, J. Sci. Comput. {\bf{34}}, 127 (2008)
\bibitem{Grima2010} R. Grima, J. Chem. Phys. {\bf{133}}, 035101 (2010)
\bibitem{Grima2011} R. Grima, P. Thomas and A. V. Straube, J. Chem. Phys. {\bf{135}}, 084103 (2011)
\bibitem{Wallace2012} E.W.J. Wallace et al., IET Sys. Biol. {\bf{6}}, 102 (2012)
\bibitem{Lestas2008} I. Lestas et al., IEEE Transactions on Automatic Control {\bf{53}}, 189 (2008)
\bibitem{Thomas2014} P. Thomas,  C. Fleck, R. Grima and N. Popovic, J. Phys. A: Math. Theor. {\bf{47}}, 455007 (2014)
\bibitem{Scott2012} M. Scott, IET Syst. Biol. {\bf{6}}, 116 (2012)
\bibitem{Thomas2013} P. Thomas, H. Matuschek and R. Grima, BMC Genomics {\bf{14(Suppl 4)}}, S5 (2013)
\bibitem{Hayot2004} F. Hayot and C. Jayaprakash, Phys. Biol. {\bf{1}}, 205 (2004)
\bibitem{Cianci2012} C. Cianci et al., Eur. Phys. J. Special Topics {\bf{212}}, 5 (2012)
\bibitem{vanKampen1976} N. G. van Kampen, Phys. Letts. {\bf{59A}}, 333 (1976)
\bibitem{Abramowitz} M. Abramowitz and I. A. Stegun, {\it Handbook of Mathematical Functions} (Dover, 1965)
\bibitem{ElfEhrenberg2003} J. Elf and M. Ehrenberg, Gen. Res. {\bf{13}}, 2475 (2003)
\bibitem{Anderson2010} D. F. Anderson, G. Craciun and T. G. Kurtz, Bull. Math. Biol. {\bf{72}}, 1947 (2010)
\bibitem{Gillespie1977} D. T. Gillespie, J. Phys. Chem. {\bf{81}}, 2340 (1977)
\bibitem{Albert} B. Alberts et al., {\it Molecular Biology of the Cell} (Garland Publishing, 1994)
\bibitem{Gillespie2013} D. T. Gillespie, A. Hellander and L. R. Petzold, J. Chem. Phys. {\bf{138}}, 170901 (2013)
\bibitem{Challenger2012} J. D. Challenger, A. J. McKane and J. Pahle, J. Stat. Mech. Theory. Exp. {\bf{11}}, 11010 (2012)
\bibitem{Thomas2010} P. Thomas,  A. V. Straube, and R. Grima, J. Chem. Phys. {\bf{133}}, 195101 (2010)
\end{thebibliography}
\end{document}